\documentclass[final,1p,times]{elsarticle} 
\usepackage{graphicx} 
\usepackage{amssymb} 
\usepackage{amsthm} 
\usepackage{lineno} 

\journal{Nuclear Physics A} 
\begin{document} 

\begin{frontmatter} 


\title{Dileptons and Direct Photons at SPS}

\author{Ruben Shahoyan}

\address{CERN, PH, CH-1211, Geneva 23, Switzerland}

\begin{abstract} 
The study of dilepton and direct photon emission was one of the main 
topics of the experimental program at the SPS devoted to the search of
signals for QGP formation. Three generations of experiments, Helios-3, 
NA38/NA50, CERES and NA60 measured $e^+e^-$ or $\mu^+\mu^-$ production 
in various colliding systems and at different energies. While lepton 
pair production in p-A collisions was found to be reasonably well 
described by the expected sources, all experiments observed in nuclear
collisions an excess 
of the yield above the extrapolation from p-A. As a result of this joint 
experimental effort we have currently a large amount of information 
characterizing this excess: its mass spectrum over the full range from 
0.2 $GeV/c^2$ up to the $J/\psi$, its transverse momentum spectra including 
their mass dependence, its angular distributions, its dependence on 
collision centrality over the complete range etc. Putting together all 
this information leads to the conclusion that what we observe is the 
long-sought thermal radiation from the fireball.
\end{abstract} 

\end{frontmatter} 



\section{Introduction}
The properties of the dilepton and real photon production in relativistic
heavy ion collisions constitute a significant fraction among the
various signatures of the Quark-Gluon Plasma (QGP) formation. Being produced
at all stages of the collision and practically lacking the final state
interactions, they convey dual information.
From one side, their production rate and kinematics are
sensitive to the properties of the surrounding matter: dominance of
the partonic or hadronic constituents, temperature, density, flow etc.
In turn, the hot and dense matter is expected to affect spectral
functions of the dilepton emission. Disentangling these two effects
poses both interpetational and experimental problems: 
the ``conventional'' sources of the observed effects need to be understood,
and the small production rates and large backgrounds require large
integrated luminosity measurements.

\section{Experimental results}
The mentioned difficulties are the reason behind the scarce results on direct
photons collected so far at the SPS. 15-20$\%$ upper limits (95$\%$ CL)
on the excess over hadronic sources ($\pi^{0}$, $\eta$ and $\eta'$
decays) were set by WA80~\cite{WA80-photons} and~\cite{CERES-photons}
in the central S-Au collisions at 200 A $GeV$. The most significant result~\cite{WA98-photons}
is the observation of up to $20\pm7 \%$ excess for $p_T>1.5~GeV/c$ in
Pb-Pb collision at 158 A $GeV$. It is well described by theoretical
models involving pQCD photons and thermal emission both from hadron
gas and QGP~\cite{GALE-photons}, but the interpretation is ambiguous: 
the thermal photon emission rate is determined
by the temperature and close to the transition point both phases provide
similar rates. 

The situation is much more advanced in the dilepton sector.
Conventionally, its mass spectrum is separated into three
regions, distingished by their dominant contributions in p-p and p-A
collisions: 
(i) Low Mass Region (LMR, $M<M_{\phi}$) where the emission is mostly determined 
by the leptonic decays of vector mesons (resonant part) and non-resonant contribution
from Dalitz decays;
(ii) Intermediate Mass Range (IMR, $M_{\phi}<M<M_{J/\psi}$) composed of
Drell-Yan dileptons and pairs coming from uncorrelated decays of
open charm (mostly $D$ and $\bar{D}$ mesons);
(iii) High Mass Range (HMR,  $M>M_{J/\psi}$) dominated by the heavy
quarkonia and Drell-Yan pairs. This part of the spectrum was covered
by \cite{ROBERTA-qm09} in this volume and will not be considered here.

\subsection{Excess in LMR and IMR}
With arrival of the ion beams on CERN SPS all dilepton experiments
observed an enhancement of the $l^{+}l^{-}$ production with
respect to the extrapolation from p-A collision.  
Helios/3~\cite{HELIOS} and NA38~\cite{NA38IMR} experiments reported the continuum excess in 
$\mu^{+}\mu^{-}$, comparing the S-W and S-U collisions at 200 A $GeV$
respectively with p-W interactions at the same energy. 

\begin{figure*}[htbp]
\centering
\resizebox{0.33\textwidth}{0.22\textheight}{%
\includegraphics*{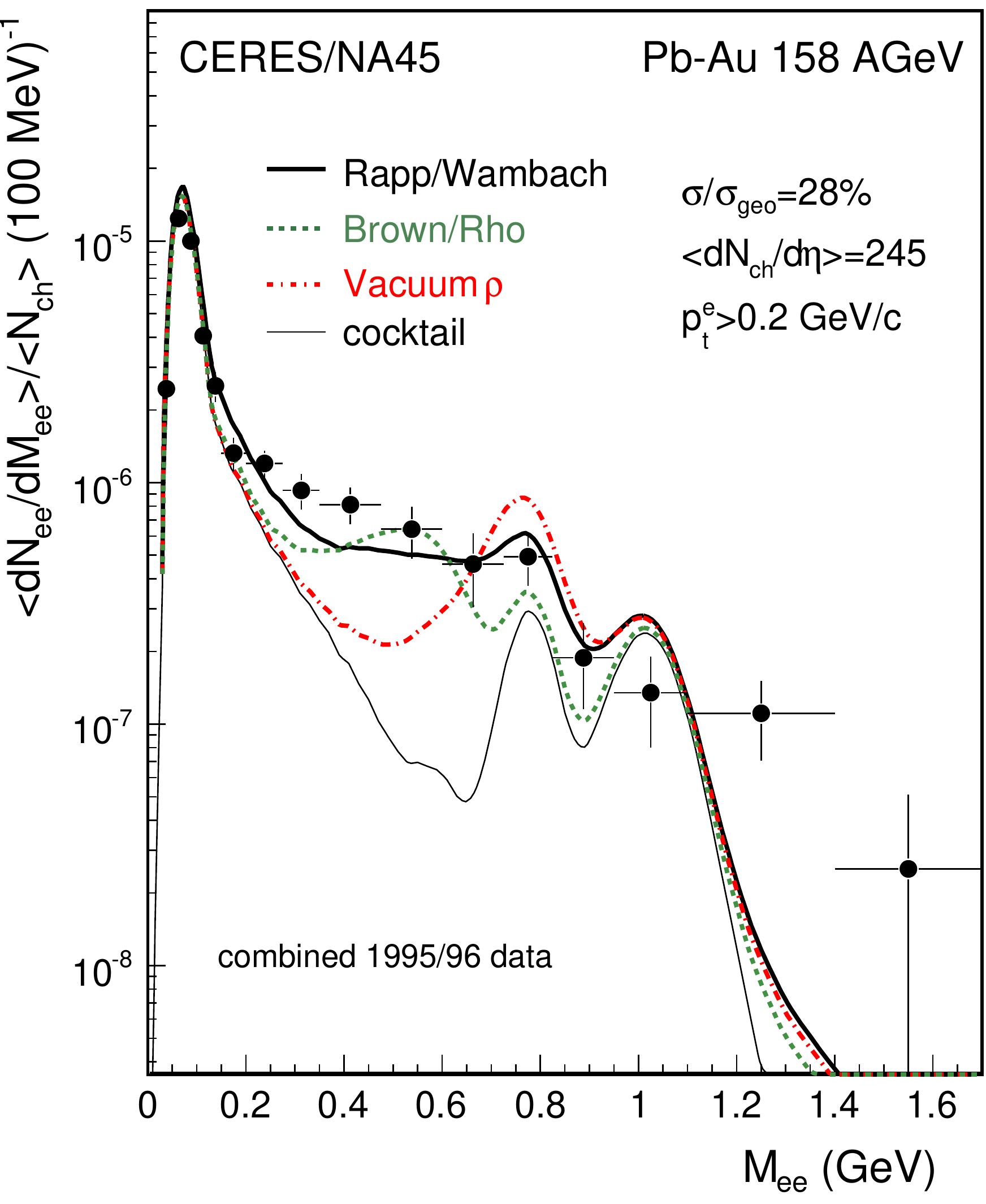}}
\resizebox{0.33\textwidth}{0.22\textheight}{%
\includegraphics*{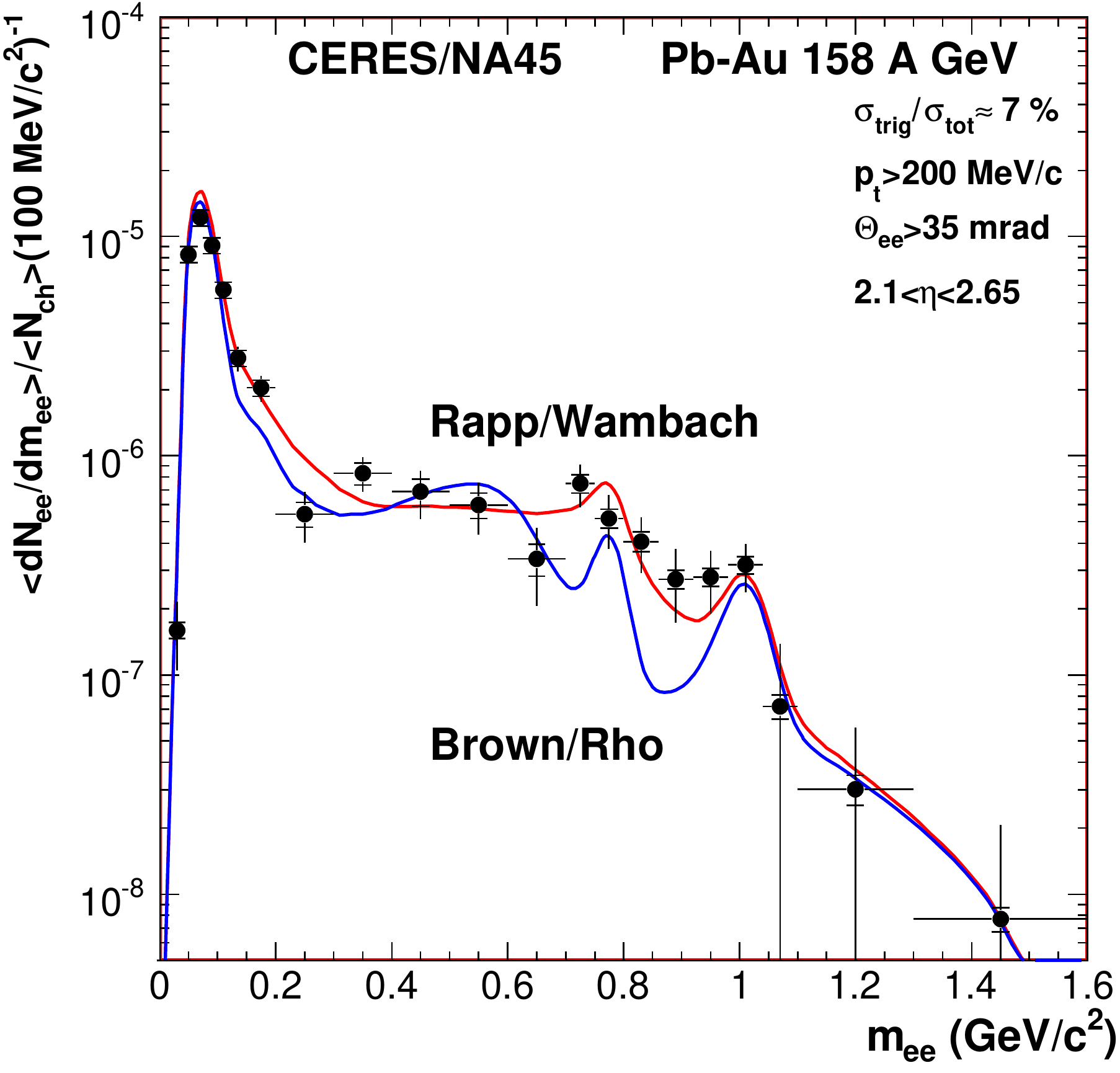}}
\resizebox{0.32\textwidth}{0.22\textheight}{%
\includegraphics*{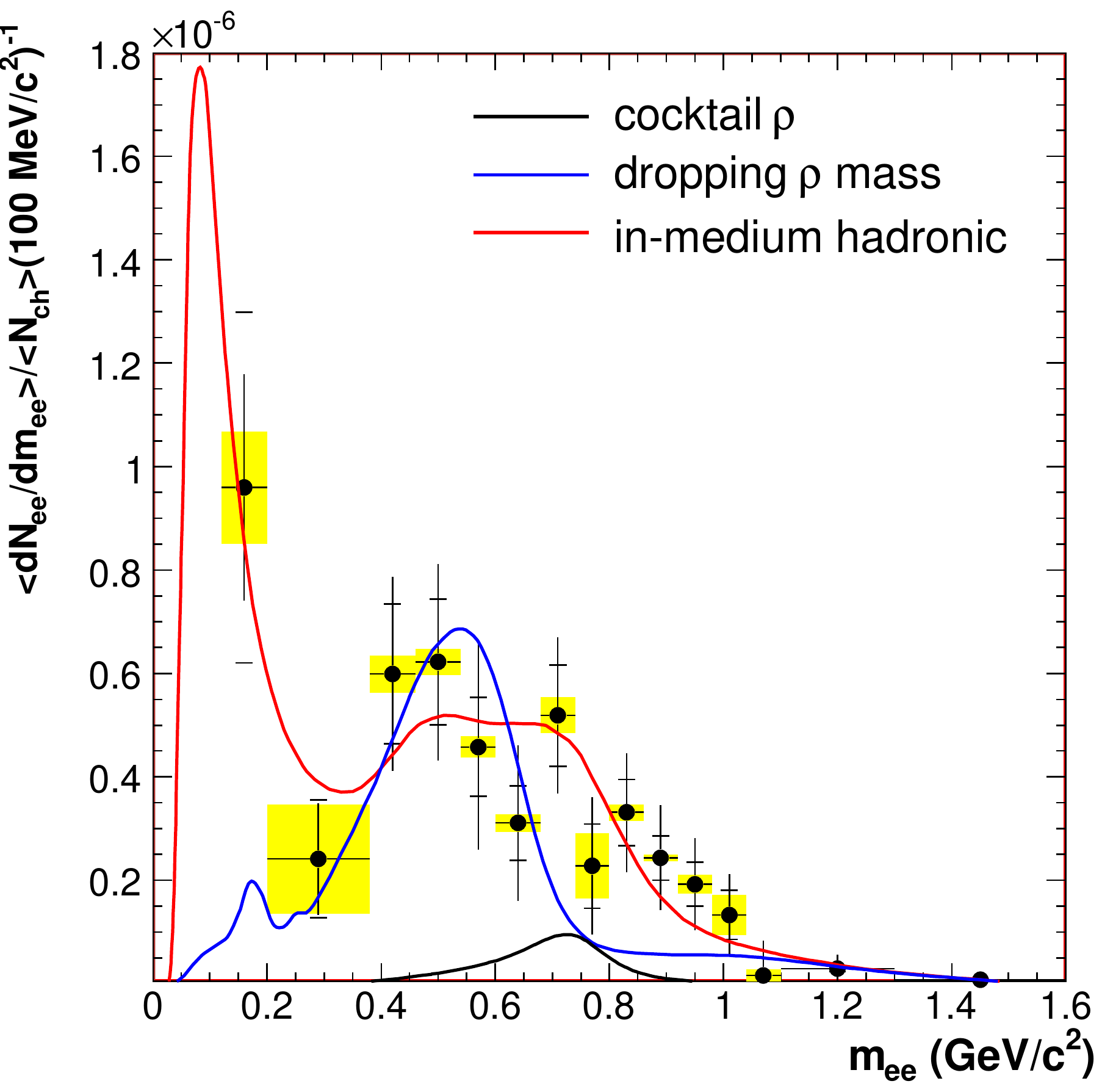}}
\vglue -3mm
\caption{
CERES inclusive $e^+e^-$ mass spectra for Pb-Au at 158 A
$GeV$ compared to models \cite{Brown-Li-Ko} and \cite{Rapp:1995}. 
\textit{(Left)}: combined 95/96 data \cite{CERES-epjc41}, 
\textit{(Center)} 2000 data,
\textit{(Right)} Same data after the subtraction of the hadronic ``cocktail''~\cite{CERES-plb666}.}
\label{fig:CERES-PbAu}
\end{figure*}

An important milestone was set by the CERES experiment which measured
electron pairs using a system of two RICH detectors. It observed
a strong ($5.\pm 0.7(stat) \pm 2.(syst)$) LMR
excess in $e^{+}e^{-}$ production in S-Au collision~\cite{CERES-prl1995} with
respect to the ``cocktail'' of hadronic decays~\cite{CERES-pBe}, well describing the
p-Be and p-Au spectra~\cite{CERES-pBe}. The triggered
theoretical activity, trying to describe these data, was not successful
until the appearence of calculations involving in-medium $\pi
\pi$ annihilation via a modified $\rho$ spectral function. The model~\cite{Brown-Li-Ko},
exploring the idea of the Brown-Rho scaling~\cite{BROWN-RHO}, assumed a 
decrease of the $\rho$ pole mass close to the chiral symmetry
restoration point, while~\cite{Rapp:1995} predicted a brodening of the
$\rho$, with significant contribution from the interactions with baryons.
Unfortunately, the low statistics and insufficient ($\sim 6\%$) mass resolution did not allow to
give preference to one of the models.
The excess was later confirmed ($2.73\pm 0.23(stat) \pm 0.65(syst)\pm
0.82(decays)$) in Pb-Au interactions at 158 A
$GeV$. Fig.\ref{fig:CERES-PbAu}\textit{(left)} shows the $e^{+}e^{-}$
combined 95/96 spectra together with the calculations
~\cite{Brown-Li-Ko} and~\cite{Rapp:1995}. 
\vglue -2mm
\vglue -2mm
After the CERES upgrade by the TPC in 1999 (which improved the mass
resolution to $\sim 4\%$), the only
existing SPS dilepton measurement at 40 A $GeV$ was
done~\cite{CERES-40GEV}. Enhancement 
$5.9\pm 1.5(stat) \pm 1.2(syst)\pm 1.8(decays)$
was reported in Pb-Au collisions, stronger than at 158 $GeV$. 
The latter was remeasured in 2000~\cite{CERES-plb666} with
the result 
$2.6\pm 0.3(stat) \pm 0.4(syst)\pm 0.8(decays)$, see Fig.~\ref{fig:CERES-PbAu}\textit{(center)}. 
Fig.~\ref{fig:CERES-PbAu}\textit{(right)} shows the isolated excess of
2000 Pb-Pb data, with the hadronic decay contributions subtracted using the rates provided by the
statistical model.
The possible explanation of the significantly larger excess at lower
energy is the  higher baryonic density due to the stronger stopping.

In the same period another important result was provided by
NA50~\cite{NA50IMR}. Using the muon spectrometer inherited from NA38 and equipped
with a very selective $\mu\mu$ trigger system, it observed in Pb-Pb
collisions at 158 A $GeV$ an IMR excess with respect to the expected 
Drell-Yan and open charm
contributions. These latter were extrapolated from the p-A
dimuon spectra at 450~$GeV$. Combined with the reanalyzed S-U data of the
NA38, the excess showed an approximately linear rise with the number of participants,
reaching a factor 2 in the central Pb-Pb collisions.
The excess, which had a kinematics resembling that of the open charm,
could be accounted for~\cite{CAPELLI} either by 
some mechanism enhancing the latter contribution in the NA50
acceptance window or by the long sought thermal dimuons~\cite{RAPPSHUR}.


The recent results are dominated by the precise and high statistics dimuon measurements of the NA60
experiment~\cite{SONDER}.
Its upgrade of the NA50 setup by a radiation
tolerant silicon pixel vertex tracker (VT)~\cite{NA60-refVT} placed
in a 2.5 Tesla dipole magnet
between the target and the hadron absorber provided numerous
advantages with respect to its predecessors.
Particularly, by matching the muons from
the spectrometer to tracks reconstructed in the VT, NA60 significantly
improved the dimuon mass resolution (from
$\sim$~80~$MeV/c^{2}$ to $\sim$20~$MeV/c^{2}$ at $\omega$ mass) and was
able to measure the offset of the muons wrt the production vertex with
a resolution of $\sim$50~$\mu m$: enough to distinguish between
prompt dimuons and those coming from the open charm decays.  

In 2003 NA60 has collected $\sim$230 millions dimuon triggers from
In-In collisions at 158 A~$GeV$. Data were taken with two current settings in the spectrometer
magnet: 4~kA for enhanced acceptance at low mass and $p_T$ and 6.5~kA
to enhance the IMR and HMR statistics. 
\vglue -2.5mm
\begin{figure*}[htbp]
\centering
\resizebox{0.32\textwidth}{0.212\textheight}{%
\includegraphics*{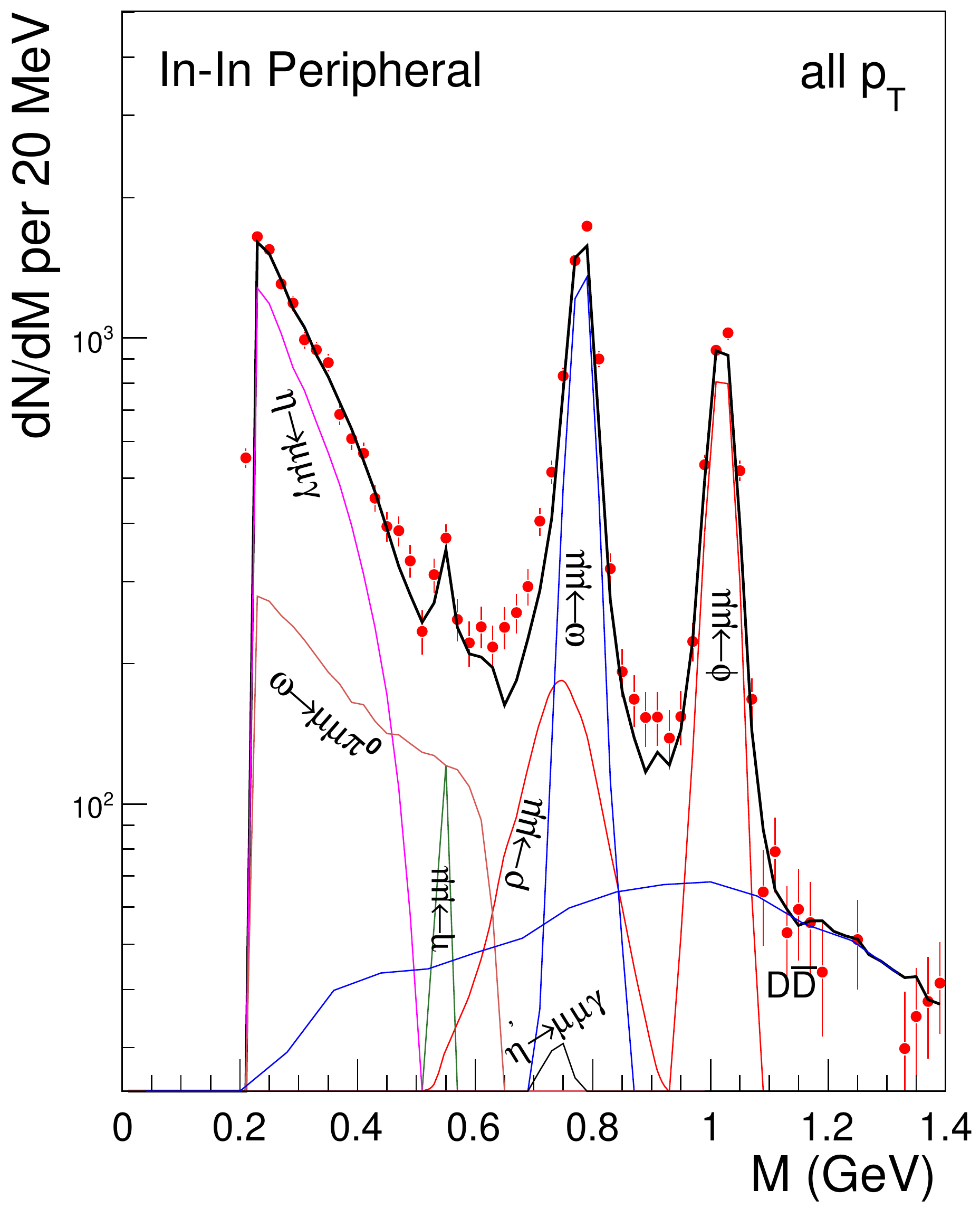}}
\resizebox{0.32\textwidth}{0.22\textheight}{%
\includegraphics*{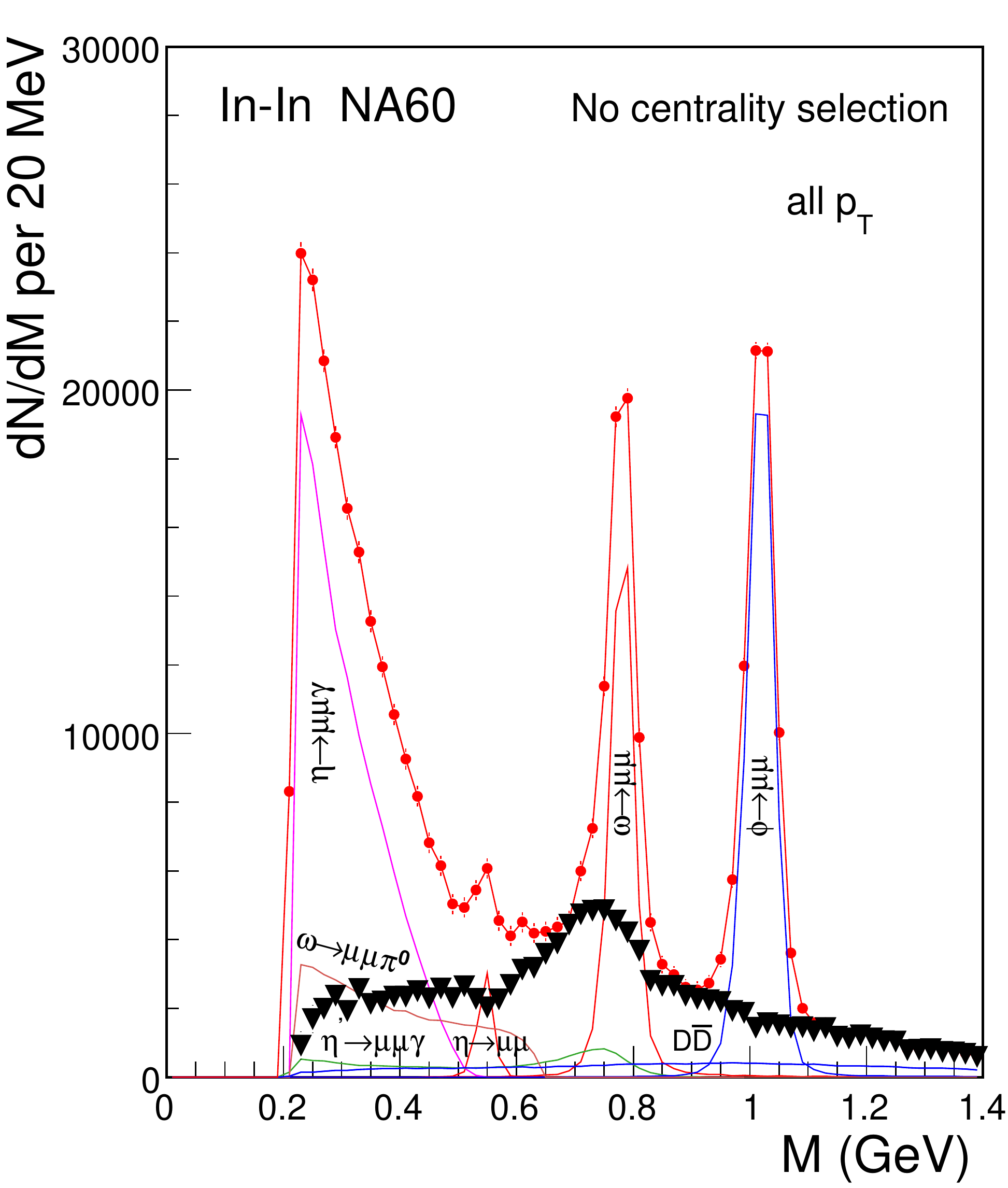}}
\resizebox{0.32\textwidth}{0.215\textheight}{%
\includegraphics*{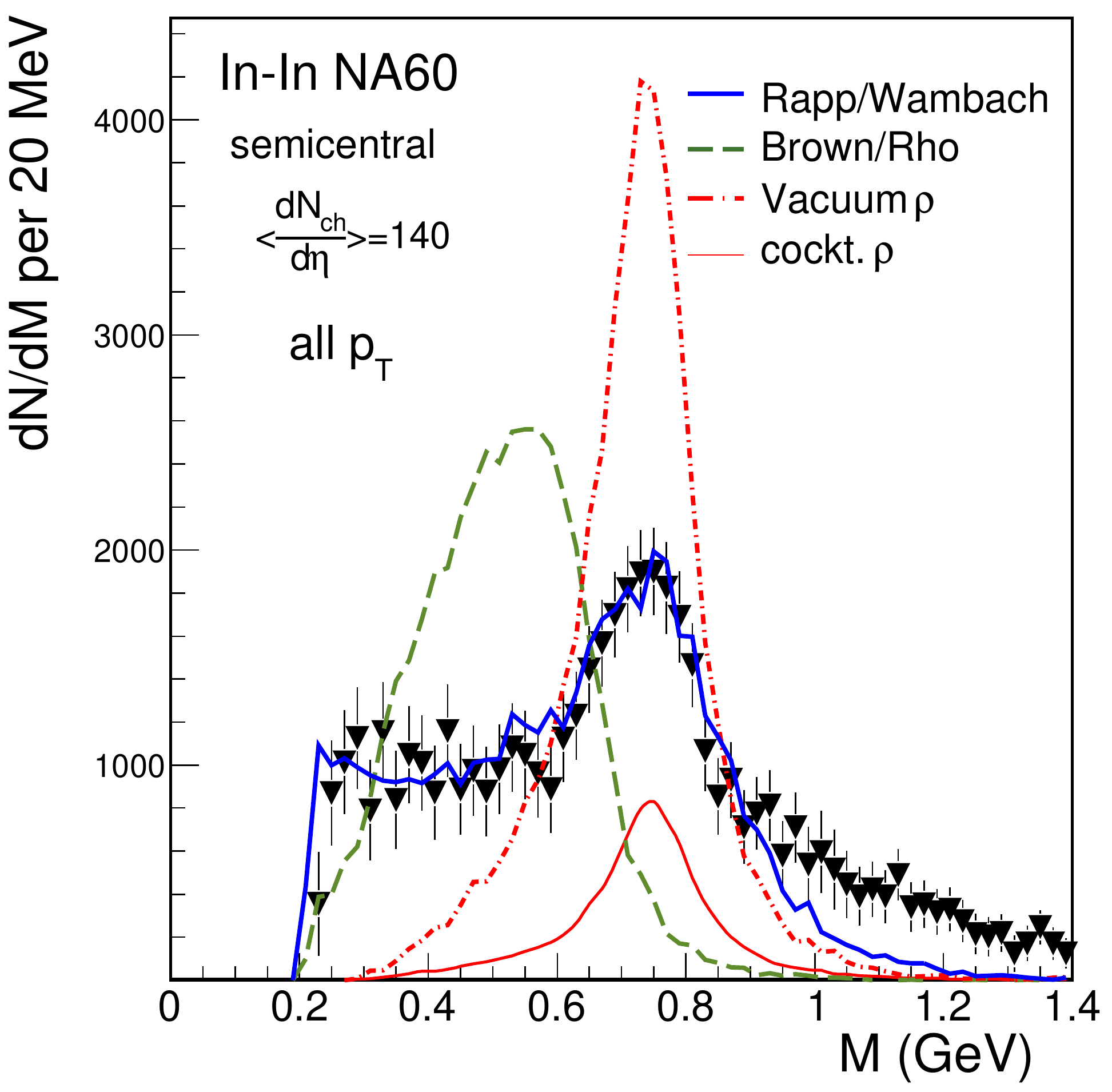}}
\vglue -2.5mm
\caption{
NA60 dimuon mass spectra measured in In-In collisions at 158 A $GeV$:
\textit{(Left)} peripheral \cite{NA60-EPJC49} 
\textit{(Center)} integrated over centrality (excess is shown by
triangles) \cite{NA60-PRL96}
\textit{(Right)} Isolated excess in semicentral collisions
\cite{NA60-PRL96} compared to the Brown-Rho and Rapp-Wambach
models~\cite{RAPP-CALC}. 
The ``cocktail'' $\rho$ is not subtracted.}
\label{fig:NA60-spect1}
\vglue -2.5mm
\end{figure*}
Fig.\ref{fig:NA60-spect1}\textit{(left)} shows the signal dimuon mass
spectrum for peripheral ($dN_{ch}/d\eta<30$) In-In
collisions~\cite{NA60-EPJC49}. It is well described by the
``cocktail'' of hadronic decays obtained using the GENESIS
generator~\cite{CERES-pBe} improved and adapted for dimuons~\cite{GENESIS-NA60}.
The high statistics and good mass resolution allowed direct fit of the
$\eta \rightarrow \mu^{+}\mu^{-}\gamma$ and $\omega \rightarrow
\mu^{+}\mu^{-}\pi^{0}$ decays form-factors~\cite{NA60-FormFactors}
with the
pole approximation $F=(1-M^{2}/\Lambda^{2})^{-1}$, yielding $\Lambda^{-2}$ (in $GeV^{-2}$)
1.95$\pm$0.17(stat.)$\pm$0.05(syst.) and 
2.24$\pm$0.06(stat.)$\pm$0.02(syst.) respectively. The values agree
with previous measurements by the Lepton-G experiment~\cite{LeptonG},
improving their errors and confirming the strong enhancement of the
$\omega$ form-factor with respect to the VMD expectation of $\Lambda^{-2}=1.68~GeV^{-2}$~\cite{VMD-FFomega}.
These measurements, together with the improved value of the 
$\omega\rightarrow\mu^{+}\mu^{-}\pi^{0}$ branching ratio ([1.73$\pm$0.25(stat.)$\pm$0.14(syst.)]$\cdot$10$^{-4}$), 
significantly decrease the uncertainty in the hadronic decay ``cocktail''
below the $\omega$ in the analysis of more central collisions.
Fig.\ref{fig:NA60-spect1}\textit{(center)} shows the LMR mass spectrum
integrated over centrality~\cite{NA60-PRL96}. Thanks to the good mass
resolution the narrow peaks of $\phi$, $\omega$ and $\eta$ as well as
the $\eta$, $\eta'$ and $\omega$ Dalitz decays can be locally
subtracted uncovering a significant excess (shown by triangles) centered
around the $\rho$ (its ``cocktail'' contribution is not subtracted). Fig.\ref{fig:NA60-spect1}\textit{(right)}
compares the isolated excess for the semi-central collisions
($dN_{ch}/d\eta=140$)~\cite{NA60-PRL96} with the preditictions of
Brown-Rho~\cite{Brown-Li-Ko} and Rapp-Wambach~\cite{Rapp:1995}
models~\cite{RAPP-CALC}. The complete disagreement with the dropping
mass and nice agreement with the broadening scenario is obvious. 
A similar conclusion~\cite{CERES-plb666},
although with less statistical siginificance, is obtained by CERES
from the Pb-Au (2000) data (Fig.~\ref{fig:CERES-PbAu}\textit{(right)}).
\vglue -2mm
\begin{figure*}[htbp]
\centering
\resizebox{0.45\textwidth}{0.2\textheight}{%
\includegraphics*{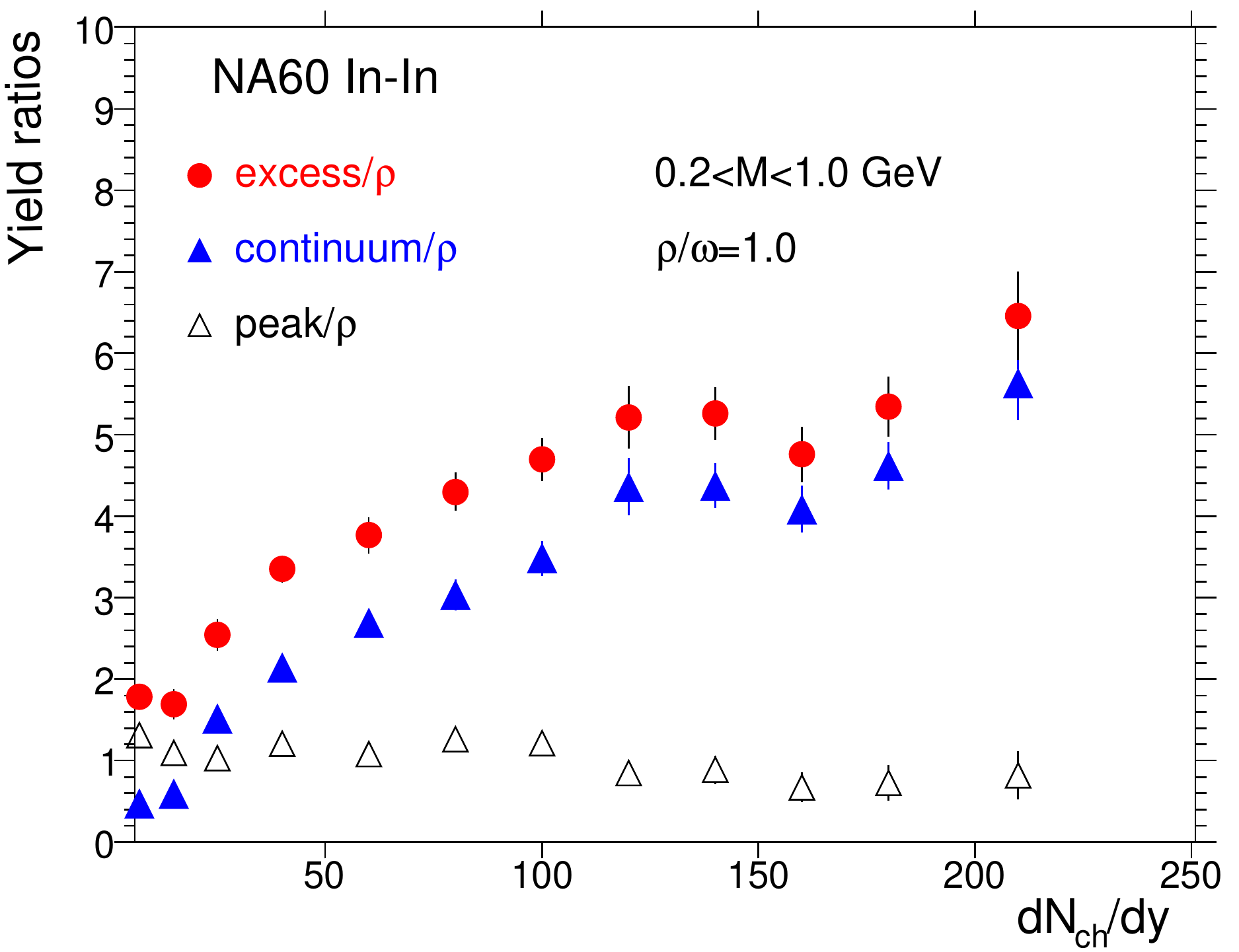}}
\resizebox{0.54\textwidth}{0.2\textheight}{%
\includegraphics*{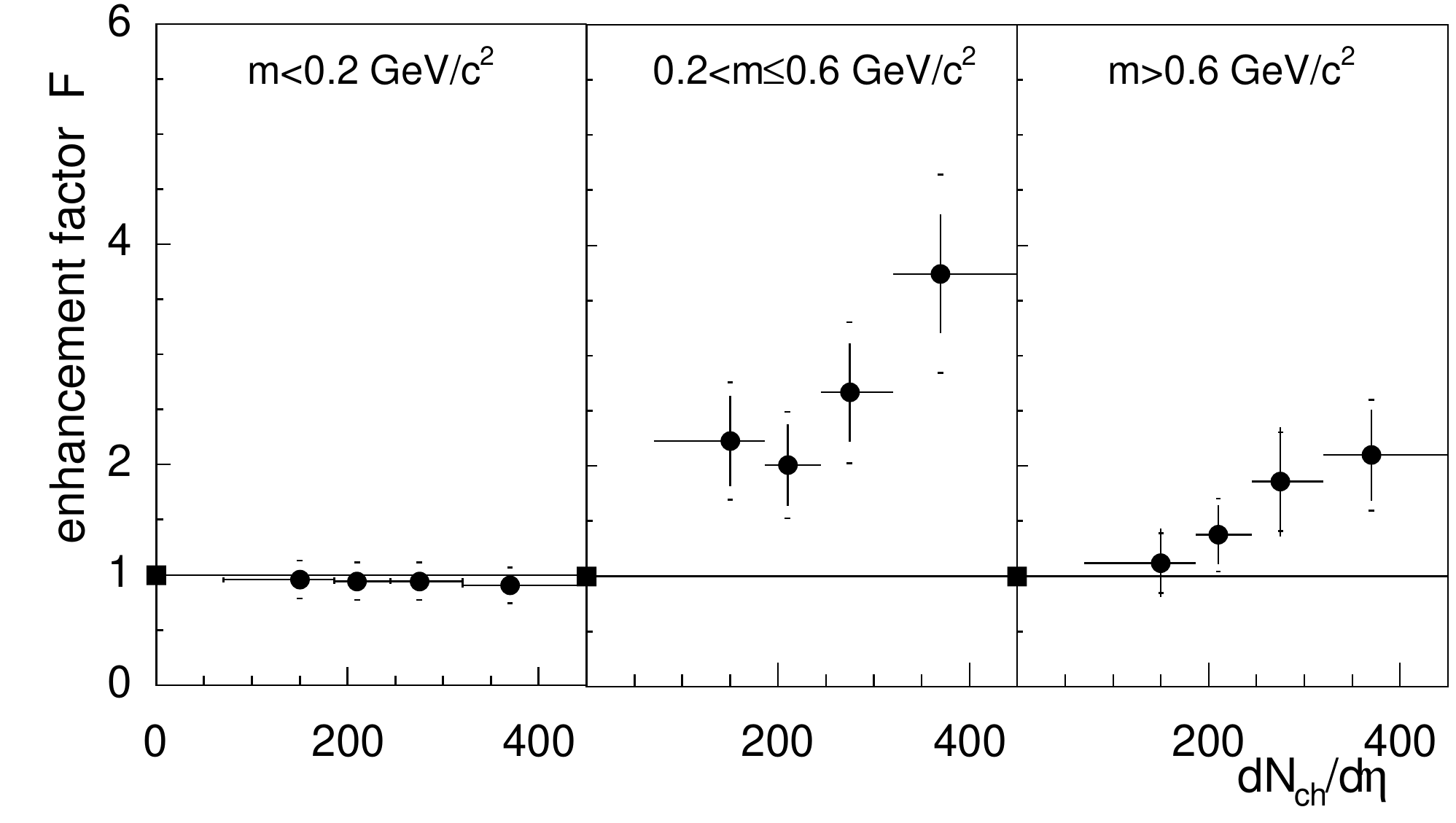}}
\vglue -2mm
\caption{Evolution of the LMR excess with the multiplicity. 
  \textit{(Left)} NA60, In-In at 158 A $GeV$\cite{NA60-EPJC49}.
\textit{(Right)} CERES, combined 95/96 Pb-Au at 158 $GeV$ data~\cite{CERES-epjc41}.
See text for details.}
\label{fig:LMR-centrality}
\vglue -2mm
\end{figure*}
Fig.~\ref{fig:LMR-centrality}\textit{(left)} shows the centrality
dependence of the LMR excess studied by NA60. The total
``excess''\textit{(circles)} is separated into continuum \textit{(filled triangles)} and
peak \textit{(hollow triangles)} parts and related to expected
``cocktail'' $\rho$ contribution (assuming $\rho/\omega=1$ on
the cross section level). While the peak contribution agrees with the
$\rho$ produced at freeze-out, the continuum part shows a
monotonic rise and broadening with centrality\cite{NA60-EPJC49}. Such a faster than
linear rise of the excess with multiplicity is compatible with emission from the
annihilation process. It can be conjectured that the magnitude of the
continuum excess directly measures the lifetime of the fireball in
number of $\rho$ generations: so called $\rho$-clock~\cite{rhoclock-heinz}.
The centrality dependence of the excess seen by CERES~\cite{CERES-epjc41} in Pb-Au
collisions, shown in Fig.~\ref{fig:LMR-centrality}\textit{(right)}
as a ratio of the excess to the total ``cocktail'' expectations, is in
a good agreement with NA60 results.

\begin{figure*}[htbp]
\centering
\resizebox{0.49\textwidth}{0.22\textheight}{%
\includegraphics*{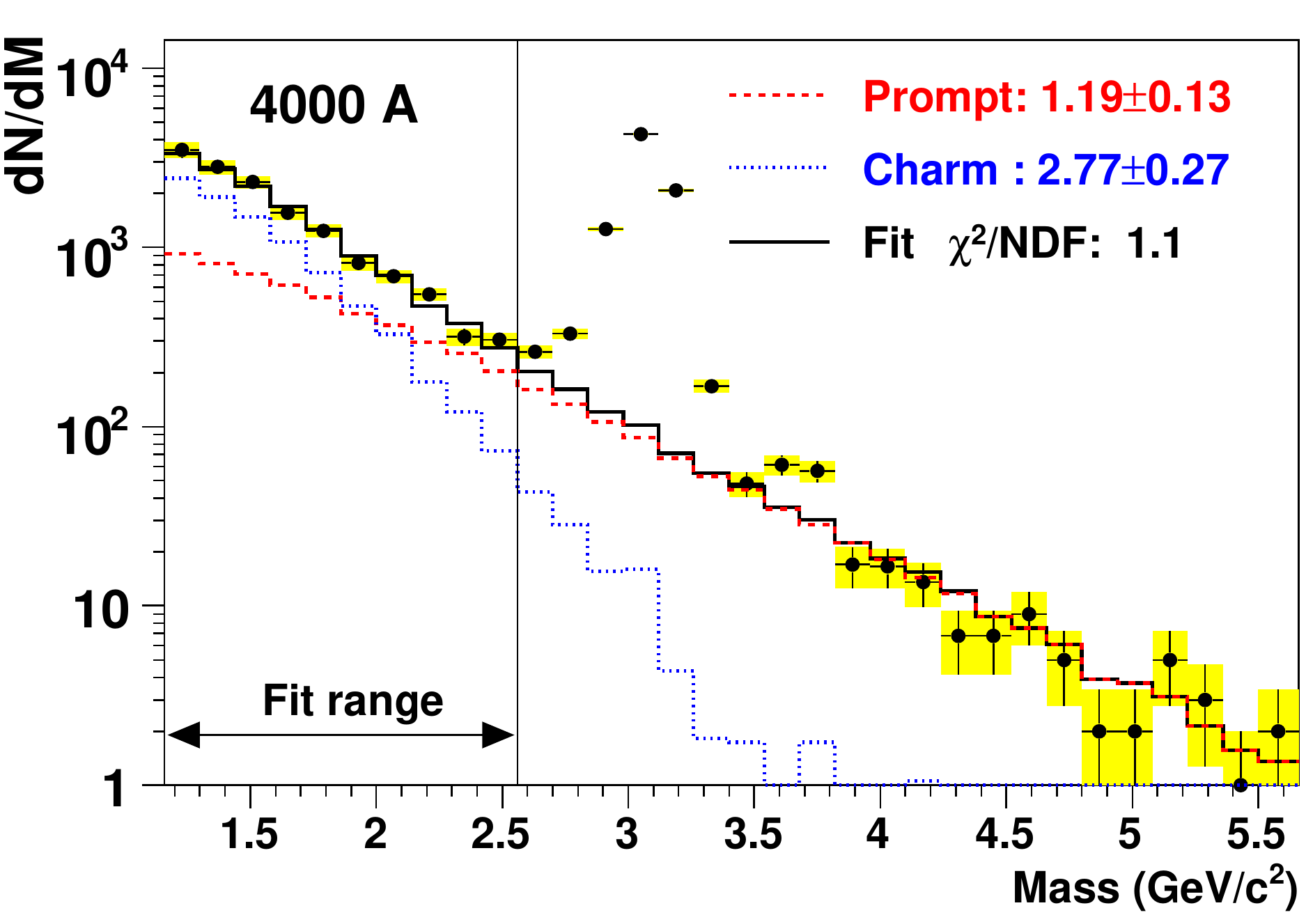}}
\resizebox{0.50\textwidth}{0.215\textheight}{%
\includegraphics*{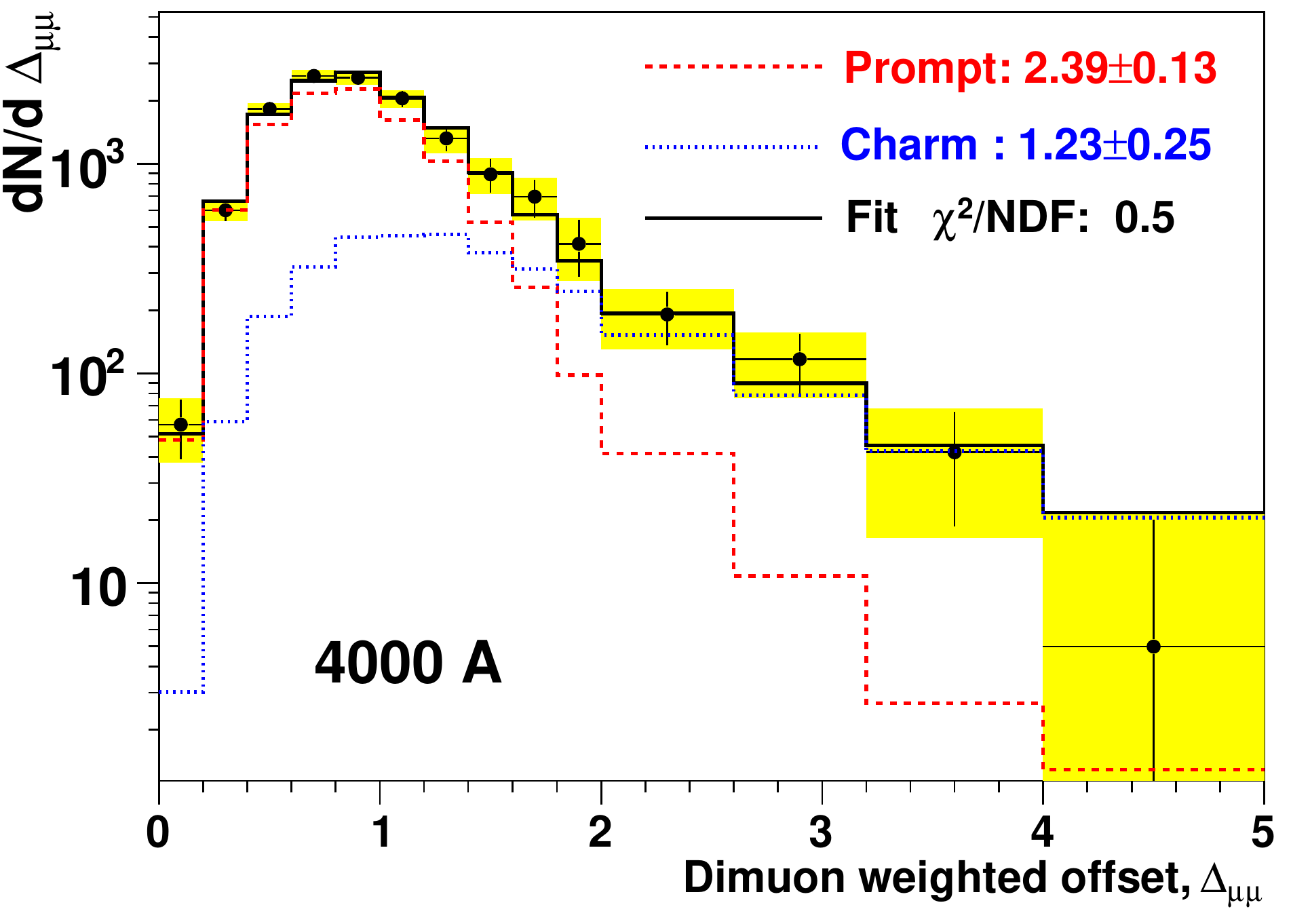}}
\vglue -2mm
\caption{IMR signal dimuon distributions in In-In
  collisions at 158 $GeV$~\cite{NA60-imr}: \textit{(Left)} mass spectra fitted in IMR by the superposition of
  Drell-Yan and open charm contributions;
\textit{(Right)} same for weighted offset spectra.
}
\label{fig:NA60-imrfits}
\end{figure*}

The analysis of the NA60 IMR mass spectrum~\cite{NA60-imr} in the In-In data is done
by fitting the signal in the $1.16<M<2.56~GeV/c^{2}$ range with a 
superposition of the Drell-Yan and open charm contributions obtained
from the Pythia~6.325 generator. The fits are done in  terms of
multiplicative factors for the reference cross sections.
The latter
are defined in the following way: for the Drell-Yan, it reproduces
the cross sections measured above the $\psi'$ by NA3~\cite{NA3DY}
and NA50~\cite{NA50DY}, while for the open charm  the
result of a similar fit to NA50 p-A data at 450 $GeV$~\cite{NA50IMR}
rescaled to 158 $GeV$ by Pythia is used 
($\sigma_{c\bar{c}} = 8.6\mu b$). Due to the insufficient high-mass
Drell-Yan statistics, the integrated effective luminosity is extracted from the number of 
$J/\psi$ events and its cross section (corrected for the nuclear and
anomalous suppression effects).  
Fig.~\ref{fig:NA60-imrfits}\textit{(left)} shows such a fit to the
centrality integrated dimuon mass spectra with low current setting in
the spectrometer. At this level the results are fully compatible with
observations of NA50 in Pb-Pb data~\cite{NA50IMR}: while the 
Drell-Yan contribution
agrees well with the expectations (and the data above $\psi'$), a
strong excess with a mass shape resembling the open charm contribution
is found. The global fit to both data sets leads to the enhancement factors
$1.26\pm0.09$ for Drell-Yan and $2.61\pm0.20$ for open charm.
To clarify the origin of this excess NA60 uses its excellent
muon offset resolution to separate (statistically) the open charm
(off-vertex decays) and prompt
contributions. Fig.~\ref{fig:NA60-imrfits}\textit{(right)} shows the
dimuon ``offset''\footnote{Defined as
$\Delta_{\mu}=({\bf d} {\bf V}^{-1} {\bf d}^{T})^{1/2}$ for the single muons
and $\Delta_{\mu \mu}=[(\Delta_{\mu 1} + \Delta_{\mu 2})/2]^{1/2}$ for
the dimuons, with ${\bf d}$ and ${\bf V}$ being the vector and
corresponding covariance matrix of the transverse
offset of the muon wrt. the primary vertex} distribution for the
mass range indicated in the left panel, fitted
to the expected open charm and prompt (assumed to be Drell-Yan) contributions.
The global fit to low and high current data sets leads to 
enhancement factor of 
$2.29\pm0.08$ and $1.16\pm0.16$ for the prompt and open charm
samples. Hence the excess should be attributed to the prompt emission,
while the open charm contribution is compatible with the extrapolation
from the NA50 pA data. 
\vglue -3mm
\vglue -2mm
The statistics of the NA60 IMR spectra is not high enough to extract the
open charm contribution differentially (in centrality, $M$,
$p_T$ etc.). For this reason, in the differential analysis the
kinematic distributions for open charm and Drell-Yan are taken from
Pythia spectra, while for the dependence on centrality both are assumed
to scale with the number of binary collisions extracted from the measured 
number of $J/\psi$ events in a given $dN_{ch}/d\eta$ bin (corrected
for the suppression). 
The IMR excess is defined as the
difference between the measured signal and the sum of these two
contributions. 

\begin{figure*}[htbp]
\centering
\resizebox{0.45\textwidth}{0.22\textheight}{%
\includegraphics*{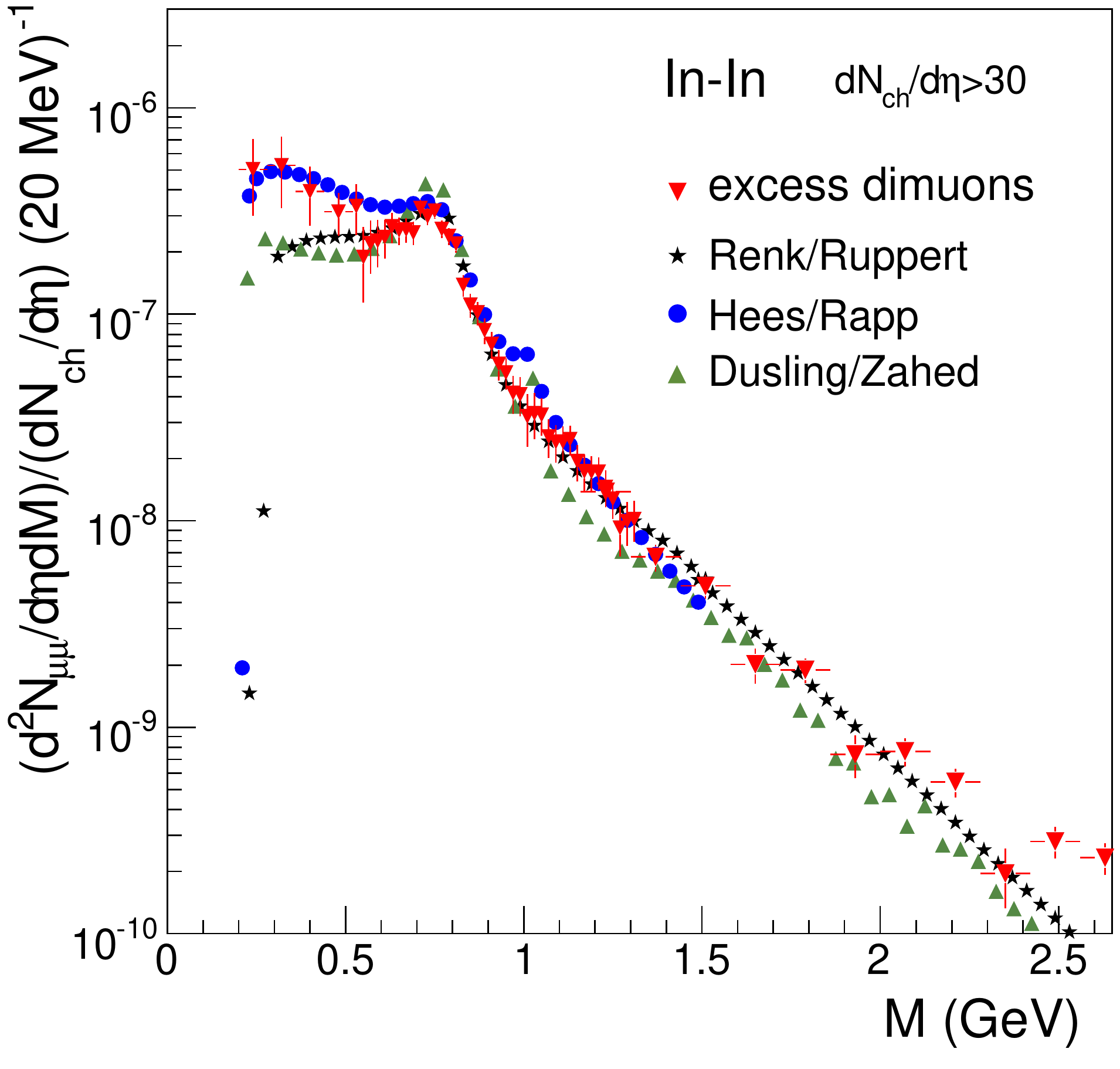}}
\resizebox{0.45\textwidth}{0.23\textheight}{%
\includegraphics*{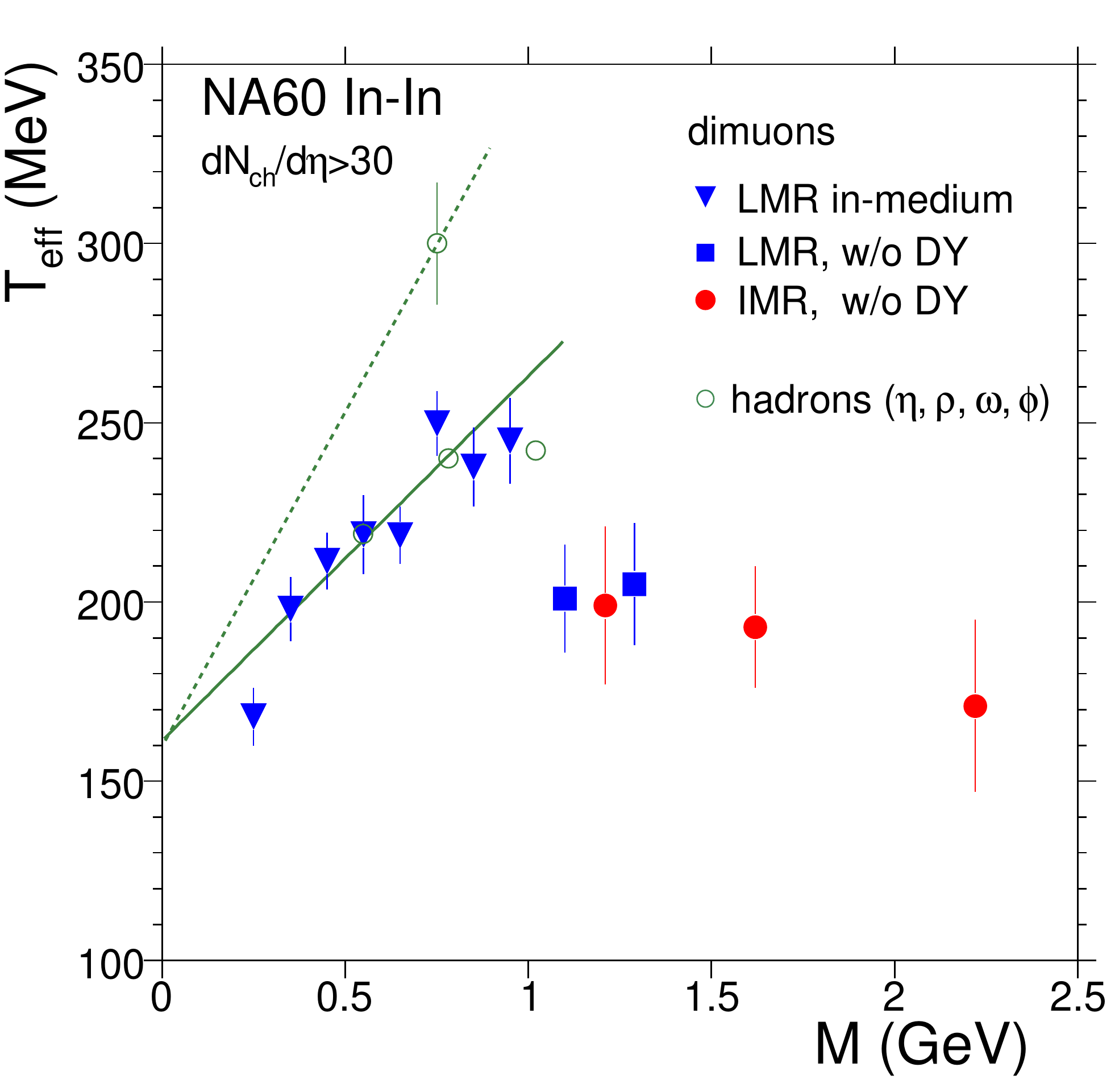}}
\vglue -3mm
\caption{\textit{(Left)}Acceptance corrected mass spectrum of the
$\mu^{+}\mu^{-}$  excess (the ``coctail'' $\rho$ is not subtracted) in In-In collisions~\cite{NA60-imr}
\textit{(Right)} T$_{\mathrm{eff}}$ of excess
(the $\rho$ peak contribution is subtracted) vs. dimuon
mass~\cite{na60-jpg-sanja,LMRna60}.
}
\label{fig:NA60-TM-comb}
\end{figure*}

The dependence of the excess on the number of the collision participants
shows a faster than linear scaling with the number of binary
collisions (with excess/DY reaching a factor $2.3\pm0.7$ for the most
central with $N_{part}>200$), but is slower than quadratic increase
with the squared number of participants\cite{NA60-imr}. As in the case of
the LMR excess such a behaviour is compatible with the emission from
the annihilation in thermalized medium.
Fig.~\ref{fig:NA60-TM-comb}\textit{(left)} summarizes the mass
spectrum of the excess seen in the In-In collisions~\cite{NA60-imr}, 
corrected for the acceptance and reconstruction efficiency and 
normalized to the per charged particle yield. 

NA60 performed also an extensive study of the $\mu \mu$ $m_T$
spectra~\cite{NA60-thermal}, summarized on Fig.~\ref{fig:NA60-TM-comb}\textit{(right)}. 
Both hadrons and excess $m_T$ spectra are
well described by the $dN/dm^{2}_{T} \propto exp{\left(-m_{T}/T_{\mathrm{eff}}\right)}$
form, except for some puzzling softening of the excess at masses
$<1.2~GeV/c^2$ in the range of $m_{T}~-~M < 0.2~GeV$. 
All hadrons - $\eta$,~$\omega$,~$\rho$ (defined as the
peak on top of the LMR excess) and $\phi$ show the rise
of $T_{\mathrm{eff}}$ with mass characteristic for the ``blue-shift''
due to radial flow. The deviations $T_{\rho} > T_{\phi}, T_{\omega}$ from
nearly linear scaling with mass ($T_{\mathrm{eff}} \approx T_{0}+M <\beta>^2$)
are compatible with different freeze-out times of the corresponding hadrons
due to their different coupling to the expanding medium. This is supported
by a ``blast-wave'' analysis~\cite{na60-jpg-sanja}: the $\phi$
decouples first, when the flow is not yet fully developed, while
the $\rho$, whith its strong coupling to pions, freezes out last
and profits from the full flow (with $T_{\mathrm{eff}}$ reaching
$\sim 300~MeV$ in the most central collisions).
A similar increase of $T_{\mathrm{eff}}$ with mass is observed for the 
LMR excess. Since, due to the ``soft point'' in the equation-of-state
\cite{SHURYAK_EOS} the 
(eventually produced) partonic phase at SPS energies is not
expected to develop significant flow, this suggest an emission from
the thermalized hadronic expanding gas. 
Surprisingly, the rise of
the excess $T_{\mathrm{eff}}$ with mass changes to the flat behaviour
after the sudden drop by nearly $50~MeV$ at $M\sim1~GeV/c^2$. The most
plausible explanation of this effect is that the excess at these
masses is dominated by thermal dimuons from a partonic source
lacking significant flow. 
\vglue -3.4mm
\vglue -3mm
Such an interpetation is supported by the 
theoretical calculations. The models labeled Hees/ Rapp \cite{HEESRAPP}
and Renk/Rupppert~\cite{RENK-RUP} assume the dominant contribution in
the LMR to be $\pi^{+}\pi^{-}$ annililation via modified $\rho$ spectral
function (\cite{Rapp:1995} and~\cite{ELETSKY} respectively), while 
the IMR excess is defined by the combination of 4 $\pi$ annihilation
processes and a significant contribution from the annihilation in the
partonic phase. They differ in the estimate of the fraction of the
latter:~\cite{HEESRAPP} puts emphasis on chiral mixing via the $\pi
a_{1}\rightarrow \mu^{+}\mu^{-}$ process, keeping the partonic contribution
within $20-60\%$ (depending on the fireball evolution scenario),
while~\cite{RENK-RUP} assumes $\sim 80\%$ contribution from the partonic phase. 
The model Dusling/Zahed~\cite{DUSLING} uses hydrodynamic calculation 
with a virial expansion for the rates in the hadronic phase and $q\bar{q}$
annihilation in the partonic one. The latter contributes $60-90\%$
to the IMR excess. All models roughly agree with the data. The
differences at low masses reflect the differences in the tail of
the $\rho$ spectrum (\cite{HEESRAPP} with a strong effect from baryons
provides the best description). A detailed comparison of these models
with data in narrow $p_T$ bins is contained in~\cite{na60-jpg-sanja}.
%

The thermal origin of the excess is further supported by the
abscence of any polarization in the excess dimuons\cite{USAI-qm09,NA60-angular}. All coefficients of the
$d\sigma / d\Omega  \propto (1+\lambda cos{^{2}\theta} + \mu sin{2\theta} cos{\phi} + \frac{\nu}{2}sin{^{2}\theta}
cos{2\phi})$ parameterization are found to be compatible with zero, 
which is a necessary (though not sufficient) condition of the 
emission from an isotropic thermalized source. 

\subsection{$\omega$ and $\phi$ mesons}
Despite various predictions~(\cite{HEESRAPP} and references therein), 
no anomaly in the pole positions and widths of the $\omega$
and $\phi$ mesons was observed so far within the experimental reach of NA60. 
Due to the longer lifetimes, only
a small fraction of these mesons produced in the medium contributes to the
observed dilepton signal. For this reason any in-medium
modifications will have much weaker effect on the measured
$\omega$ and $\phi$ spectra than for the $\rho$. Besides that, 
the possible broadening of the $\omega$ would be practically 
inobservable since it would merge with the $\rho$ peak. 
Instead one could
look for the deficit of low-$p_{T}$ $\omega$ dileptons (with
strongest contribution from in-medium decays) in the nominal pole
position. NA60 has reported the first observation of such an
effect~\cite{NA60-thermal} for the $\omega$. While the $p_{T}$ spectra
of the $\phi$ in the whole $p_T$ range and of the $\omega$ at
$p_{T}>0.8~GeV/c$ are perfectly thermal and  
agree very well with the ``blast-wave'' fits at all centralities, the
low-$p_T$ $\omega$ spectra become gradually depleted as centrality
increases, with almost complete disappearence of the $\omega$ with
$p_{T}<0.2~GeV/c$ for the most central collisions. 

\begin{figure*}[htbp]
\centering
\resizebox{0.46\textwidth}{0.23\textheight}{%
\includegraphics*{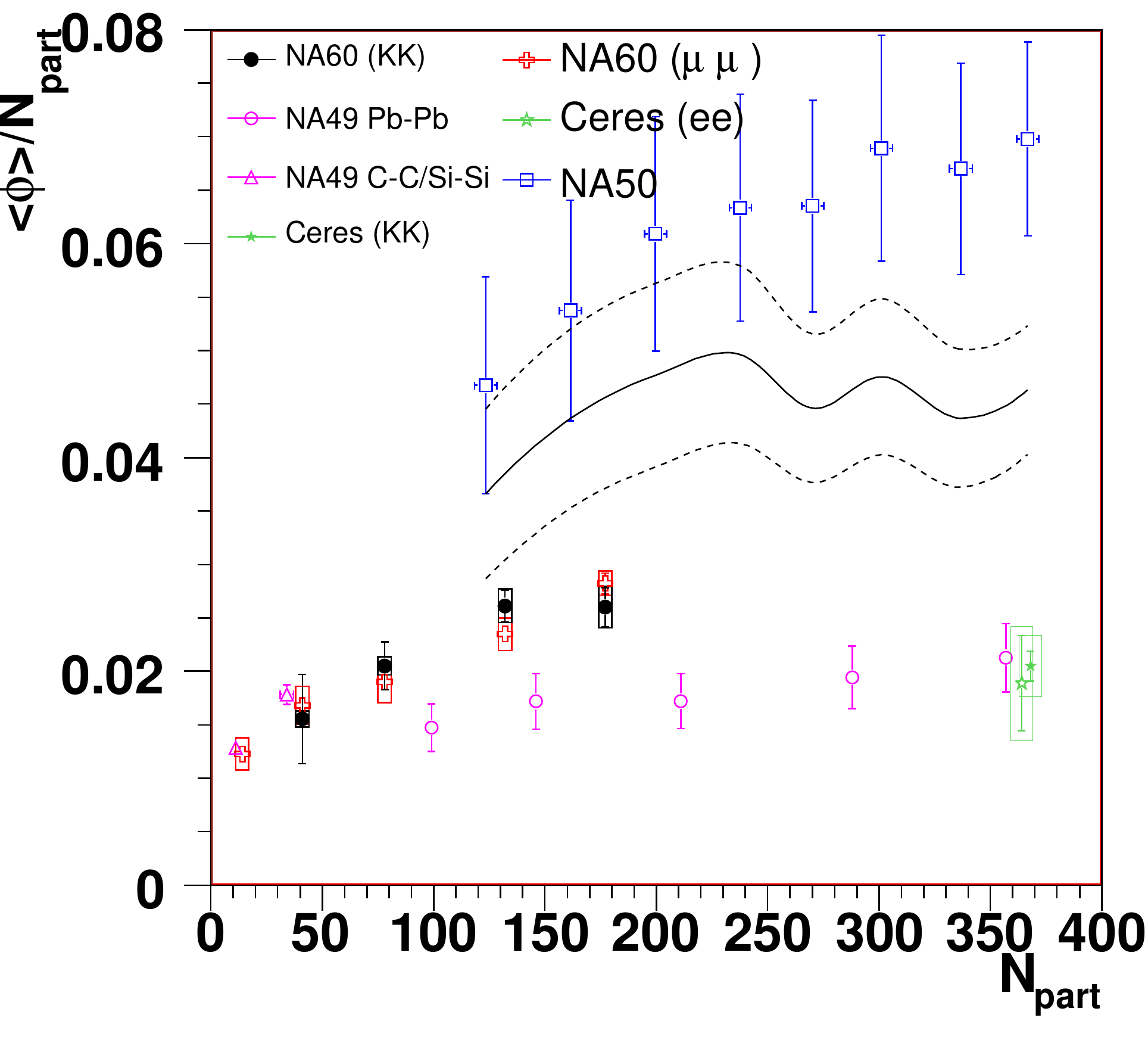}}
\resizebox{0.46\textwidth}{0.23\textheight}{%
\includegraphics*{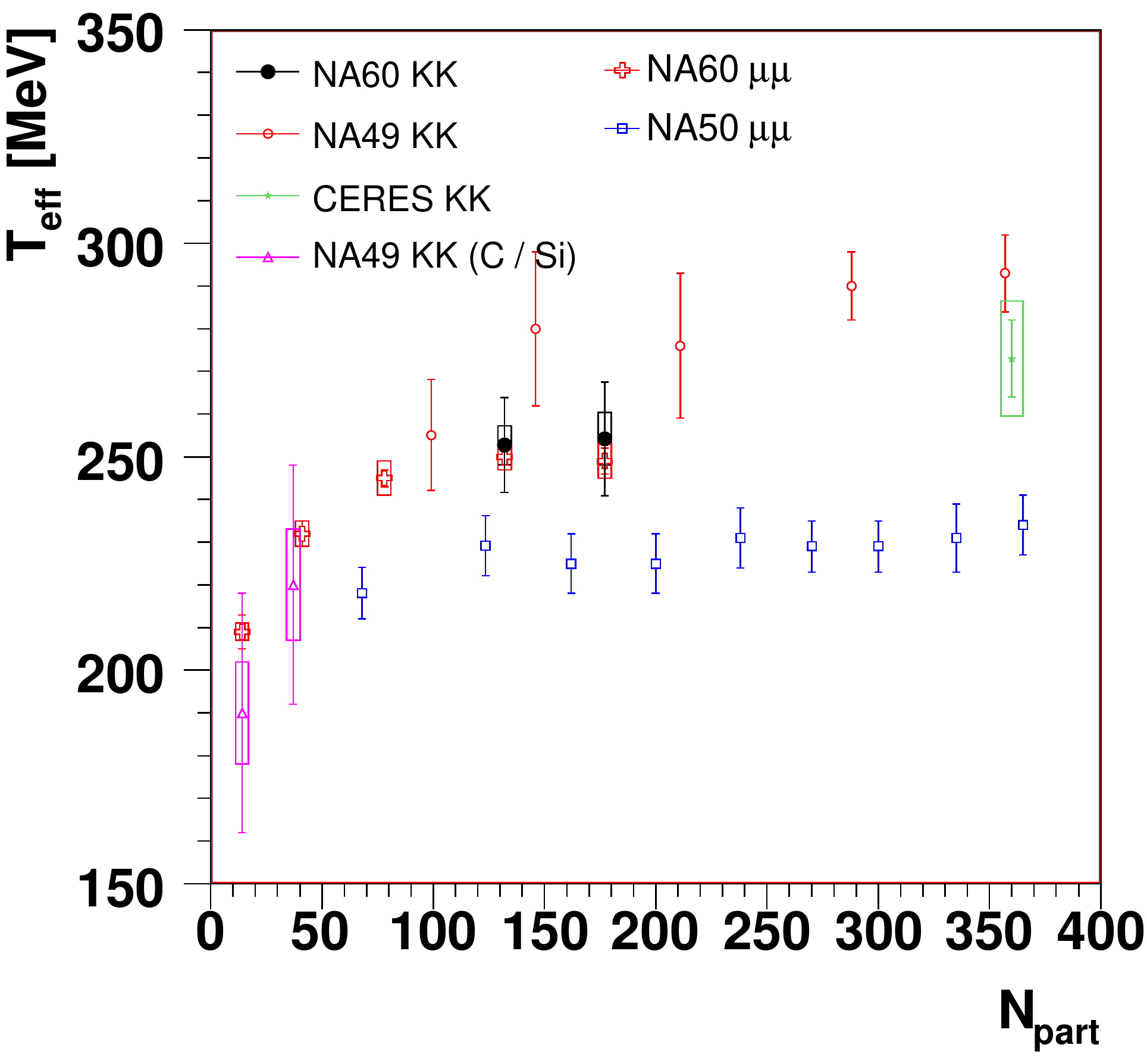}}
\vglue -4mm
\caption{Mean $\phi$ multipliciy per participant~\textit{(Left)} and  
  $T_{\mathrm{eff}}$ \textit{(Right)} as a function of the number of
   participants~\cite{ADF-qm09}.
}
\vglue -4mm
\label{fig:phi}
\end{figure*}

Significant attention was paid recently to so called $\phi$-puzzle:
the disagreement between $\phi\rightarrow K^{+} K^{-}$ measured 
by NA49~\cite{NA49-phi} and $\phi\rightarrow \mu^{+} \mu^{-}$ measured by
NA50~\cite{NA50-phi}. 
NA50, whose $\phi$ acceptance is limited by
$p_{T}>1.1~GeV/c$ sees nearly twice the yield observed by NA49 whose
statistics is limited by $p_{T}<1.6~GeV/c$, with a significant difference
in the inverse slope parameters:
$T_{\mu \mu} = 234\pm 7$ vs. $T_{KK} = 305\pm 15~MeV$ in the central Pb-Pb
collisions. Recent
measurements by CERES~\cite{CERES-phi} of $\phi$ production in
central Pb-Au collisions at 158 A $GeV$ both in the $K^{+}K^{-}$ 
and $e^{+}e^{-}$ channels 
are consistent with each other and seem to confirm the NA49 results
(although with large errors on the $T_{\mathrm{eff}}$ in
$e^{+}e^{-}$). NA60 has reported similar measurements of both
$K^{+}K^{-}$ and  $\mu^{+} \mu^{-}$ channels in In-In collisions~\cite{ADF-qm09}.
Like CERES, it finds a good agreement between the two channels, with
$T_{\mathrm{eff}}$(Fig.~\ref{fig:phi}\textit{(right)}) compatible with the observations of NA49 for the same
number of participants, but it observes a
smaller $\phi$ (Fig.~\ref{fig:phi}\textit{(left)}) yield per
participant than NA50 and slighly stronger than both NA49 and CERES
do.
Taking into account that NA50 recently reanalysed its Pb-Pb data and confirmed the 
previous results~\cite{NA50DJ-phi}, at present, it is difficult to reconcile all of the observations into
a coherent picture, albeit there is some hint for a possible physics
mechanism leading to a difference in the two channels~\cite{NA60-phi}.
\vglue -3mm

\section{Conclusions}

A summary of dilepton measurements at the SPS was
presented. The most prominent result is the excess observed in heavy
ion collisions at all masses below the $J/\psi$. Its most plausible
explanation by the production of thermal dimuons is supported by
the thermal-like spectra both in mass and in transverse momentum, 
the lack
of any polarisation and by the reasonable agreement with theory. At
low masses, the $l^+l^-$ pairs are dominated by $\pi \pi$ annihilation
via the short-lived $\rho$ meson broadened by the hot medium (the Brown-Rho scenario
of a dropping $\rho$ mass is ruled out by the data). The monotonic rise of
$T_{\mathrm{eff}}$ with mass up to $M\sim 1~GeV/c^2$ with a sudden
drop and then stabilization at higher masses, seen by NA60,
suggests that in the IMR the dominant contribution is due to the
$q\bar{q}$ anniliation in the partonic phase, when the radial flow
has not yet developed.
The suppression of the low-$p_T$ $\omega$ mesons in the central In-In
collisions hints on the first observation of its in-medium
modifications. The $\phi$-puzzle: the contradiction between the $l^+l^-$
 and $K^+K^-$ decay channels studied by the NA50 and NA49 experiments
in Pb-Pb collisions is not solved despite 
the consistent results
obtained for both channels by CERES and NA60 in Pb-Au and In-In
collisions, respectively.
\vglue -2mm

\section{Acknowledgments}
The author is grateful to the organizers of the Quark Matter 2009
conference for financial support of his participation.

\end{document}